\documentstyle[twocolumn,prl,aps]{revtex}
\begin{document}
\draft
\preprint{}
\title{Spin-Peierls vs.\ Peierls distortions \\
in a family of conjugated polymers}

\author{ M.A. Garcia-Bach$^1$,
         R. Valent\'{\i}$^2$,
         D.J. Klein$^3$}

\address {$^1$ Departament de F\'{\i}sica Fonamental,
Facultat de F\'{\i}sica, Universitat de Barcelona, Av.\ Diagonal,
647, 08028 Barcelona, Catalunya, Spain. \\}

\address{$^2$ Institut f\"ur Physik, Universit\"at Dortmund, 44221
Dortmund, Germany. \\}

\address{$^3$ Texas A \& M University at Galveston, Galveston,
Texas 77553-1675.}

\date{\today}

\maketitle

\begin{abstract}

\hspace{\parindent}  Distortions in a family of conjugated polymers
are studied within two complementary approaches, i.e.\ within a
many-body Valence Bond (VB) approach using a transfer matrix
technique to treat the Heisenberg model of the systems, and also in
terms of the tight-binding band-theoretic model with interactions
limited to nearest neighbors.  The computations indicate that both
methods predict the presence or absence of the same distortions in
most of the polymers studied.
\end{abstract}


\newpage
\section{Introduction}

The recent discovery of the first inorganic spin-Peierls material,
CuGeO$_3$, \cite{Hase} has engendered a renewed interest in
Spin-Peierls systems, i.e.\ systems which present a structural
distortion below the spin-Peierls temperature due to residual
magnetoelastic couplings stabilizing the ground-state, in analogy
to Peierls distortion \cite{Peierls} associated to an electron-soft
phonon instability opening a band gap at the Fermi level.  Recent
experiments \cite{Weiden} suggest that this is not an isolated
case, and the pronounced decrease of susceptibility observed
\cite{Isobe} in $\alpha$-NaV$_2$O$_5$ is also due to a spin-Peierls
transition.

The spin-Peierls transition was first observed in predominantly
organic compounds as TTFCuBDT \cite{Bray75}, TTF-TCNQ \cite{Andre},
(TMTSF)$_2$PF$_6$ \cite{Bechgaard}, or TTF-AuBDT \cite{Bray85}.
Theoretically, it has been studied (see for instance
\cite{Beni,Bulaesvskii,K-A79,Cross,Kondo,Nakano} and Refs.\ therein)
as a geometrical symmetry breaking for the lowest
eigenstate of a Heisenberg Hamiltonian.  Peierls and spin-Peierls
phenomena are still a subject of discussion for many other
polymers, since if a deviation occurs that lowers the chain's
symmetry, then different symmetry-equivalent distorted
ground-states may arise which correspond to different thermodynamic
phases and, at sufficiently low temperature, the possibility of
solitonic excitations and/or conduction could arise \cite{Seitz}.

Furthermore, it has been argued \cite{Klein86} that under similar
structural circumstances a Peierls distortion is predicted for the
simple H\"uckel tight-binding model of $\pi$-network strips if and
only if a spin-Peierls distortion is also predicted from Valence
Bond (VB) theory (or the formally equivalent $s=1/2$ Heisenberg
model) at the simple resonance theoretic level.  At this level of
approximation the VB wave-functions are restricted to equally
weighted superpositions of special covalent VB singlet states,
i.e.\ of Kekul\'e structures \cite{Pauling}, where every
$\pi$-electron is coupled to a singlet state with one of their
nearest-neighbors.  These Kekul\'e structures may be partitioned
into long-range ordered spin-pairing phases, the lowest-lying
phase corresponding to the highest count of Kekul\'e structures
contributing to it.  Within this approach, a spin-Peierls
distortion is predicted if there are two
maximum-cardinality-degenerate Kekul\'e phases (see \cite{Klein91}
and references therein).  Then, this
correspondence between Peierls and spin-Peierls instabilities
implies that a zero-width band gap for a $\pi$-network polymer is
predicted if and only if there are two such cardinality-degenerate
Kekul\'e phases.  The question then arises as to whether this
correspondence is maintained when going beyond the resonance
theoretic approximation.

For instance, the dimerization in polyacetylene has traditionally
been interpreted in terms of band theory \cite{Longuet,Su} as a
Peierls distortion.  Recently, however, this dimerization has also
been successfully explained \cite{pa} with a Heisenberg-like
Hamiltonian model \cite{JACS} as a spin-Peierls distortion, using
both cluster-expanded wave-functions and perturbation theory.  This
cluster-expanded many-body treatment of distortions has also been
applied to the polyacene polymer \cite{pace}, which earlier has
been extensively studied from the independent particle point of
view, since it exhibits an accidental zero-width band gap at a
simple tight-binding level (see \cite{pace,gao} and references
therein), and a new quasi-degeneracy has been predicted.

Thence, the comparison between the independent-particle and the
many-body VB treatments for degeneracy and symmetry-breaking in
polymers deserves further analysis.  It is our purpose here
to investigate the ground-state symmetries and degeneracies for
several conjugated polymers using both a simple many-body VB
framework and a simple tight-binding model.  The rationale for
these simplest models (with just nearest-neighbor interactions) is
that they reveal distortive responses which qualitative dominate
over the otherwise harmonic responses (e.g., associated with the
$\sigma$ electrons.  That is, these simplest models should reveal
dominant qualitative features, which should persist independently
of parameterization.

The polymers we focus our attention on are:  polyaceacene (PAA),
poly\-(benz[m,n])\-anthracene (PBA), and polyperylene (PPR)
(Fig.\ \ref{fig:polymers}).  All these systems exhibit a zero-width
band gap at the simplest tight-binding level.  So far, very few
experimental results are available.  Only PBA \cite{Tanaka-PBA}, and
PPR \cite{Murakami} have already been synthesized.
Theoretically PPR has been treated from the independent particle
point of view \cite{Tanaka-PPR,Bozovic}, and also using the valence
effective hamiltonian technique \cite{Bredas}, while as far as we know
PBA has not been previously treated.  PAA has been discussed in the
literature, mostly from an independent particle point of view
\cite{Bozovic,Yamabe,Kertesz,Tanaka-87,Klein-94} and less
frequently from a resonance-theoretic approach \cite{Seitz,Klein86},
though it has not been synthesized yet.  PAA can be seen, together
with polyacetylene and polyacene, as the first members of a family
of poly-trans-polyacetylenes, graphite being the final member of the
family.  All these can be thought of as special cases of ladder
materials \cite{ladder} as already had been pointed out in
Refs.\ \cite{Seitz,Tanaka-87,Klein-94}.

Within the many-body VB framework, we will consider the
antiferromagnetically-signed spin-$\frac{1}{2}$ Heisenberg model
(for more general derivations of this model than those based on
degenerate perturbation expansions see for instance
\cite{JACS,Heisenberg} and references therein).  Adequate many-body
wave-function {\it ans\"atze\/} provide variational upper bounds to
the ground-state energy.  Two different kinds of variational
localized-site cluster expanded {\it ans\"atze\/} have been
considered:  first a {\it Resonating\/} VB (RVB) {\it ansatz\/},
where the trial wave-function is a weighted superposition over all
singlets constructed as products of singlet pairs each involving
two (not necessarily nearest-neighbor) sites at a time; and second
a {\it N\'{e}el-state-based ansatz\/}, where a N\'{e}el state is
the zeroth order wave-function from which the trial wave-function
is generated.  We evaluate the matrix elements for each {\it
ansatz\/} with a transfer-matrix technique introduced previously
\cite{pa,pace,Klein86-2,Mod,teochem,croatica}.  For the
tight-binding band theory calculations we consider the so-called
translationally adapted H\"uckel model limited to nearest
neighbors.

This paper is organized as follows:  in section \ref{sec:polymers}
the description of the polymers, their symmetries and relevant
distortions are given.  In section \ref{sec:huckel} we introduce
briefly the translationally adapted H\"uckel model.  In section
\ref{sec:VB} a description of the VB method is given in terms of
the Heisenberg Hamiltonian and the trial wave-functions are
presented.  Also, the technique to compute the physical magnitudes
based on a transfer matrix is introduced and applied to obtain 
the ground-state energy of the systems.  Results are presented and
discussed in section \ref{sec:resultats}.  And finally our
conclusions can be found in section \ref{sec:conclusions}.

\section{Description of the polymers and their symmetries}
\label{sec:polymers}

The systems studied are polymeric strips of finite width and infinite
length ($L \rightarrow \infty$) (see Fig.\ \ref{fig:polymers}).  They
are constructed with fused benzene rings, and can be seen as cut from
the 2-dimensional graphite or honeycomb lattice.  Each site of the
lattice is taken to represent an $sp^2$-hybridized carbon atom with
one $\pi$-orbital perpendicular to the plane of the lattice and with
one $\pi$-electron per site.  These strips are presumed to be {\it
translationally symmetric\/} along $L$, with periodic boundary
conditions; so that the strips may be divided into {\it unit
cells\/} or eventually {\it reduced unit cells\/}, when the {\it
space group\/} of the strip contains operations involving glide
reflections.  The {\it space group\/} of the strips include, along
with the primitive translation, rotations $C_n$, reflections
$\sigma$, and combination of rotations and reflections (improper
rotations), coordinate inversion $i$, and screw-rotations and
glide-reflections, $C_s$, i.e.\ a combination of an improper
two-fold rotation or reflection with a nonprimitive translation of
half a unit cell which by themselves do not leave the lattice
invariant (see for instance Fig.\ \ref{fig:symmetries} and Table
\ref{tab:grups}).

Of special interest are minimal subsets of symmetry operations,
whose removal lead to:  (i) a band gap opening at the Fermi level,
when analyzed from the band-theoretic point of view;  (ii) the
lifting of the degeneracy of Kekul\'e phases, if seen from the
resonance-theoretic treatment.  If a zero gap occurs at $k=\pi$
(as is frequently the case for benzenoid polymers) then such a
minimal subset will be so as to no more than double the size of a
unit cell.

When a symmetry is broken, there is a distortion parameter
$\Delta_l$ associated to the stretching or shortening of the bond
$l$, with bonds numbered as in Figs.\ \ref{fig:PAA}, \ref{fig:PBA}
and \ref{fig:PPR}.  Two symmetry elements are chosen to label the
interesting distortions for every polymer as shown in Tables
\ref{tab:PAA}, \ref{tab:PBA} and \ref{tab:PPR}, where appropriate
constraints on the $\Delta_l$, imposed by the different symmetry
breakings, are also shown.  The distortions are classified as to
symmetric (+1) or antisymmetric (-1) with respect to these two
selected symmetry elements.

The PAA polymer is formed by benzene rings sharing four consecutive
edges with neighboring rings as shown in Fig.\ \ref{fig:polymers}(a).
It can also be seen as a trimer of non-dimerized parallel all-trans
polyacetylene chains.  The 6-site unit cell can be broken into two
3-site reduced unit cells, defined as the region between dashed lines
in Fig.\ \ref{fig:polymers}(a).  In the band picture, there is a
half-filled band and, consequently, a {\it zero-width band gap\/}
is predicted, regardless of distortions which preserve the
glide-reflection symmetry.  In the simplest VB picture,
i.e.\ resonance theory, there are two maximum-cardinality
degenerate Kekul\'e phases.  For instance, defining $M$
as the number of ``double bonds'' crossed by an oblique line (see
Fig.\ \ref{fig:phases}), there are two Kekul\'e phases $M$= even
equivalent to two $M$= odd which do not mix because of the cyclic
boundary conditions of the strip and they are degenerate since they
each contain essentially a single Kekul\'e structure.  A distortion
that could open the band gap at the Fermi level and lift the
degeneracy of the Kekul\'e phases requires the destruction of the
glide-reflection symmetry.  The distortions to be considered are
then those which are {\it antisymmetric\/} with respect to
interchange of the two reduced unit cells in a new unit cell.

PBA is formed by a polyacene strip where added benzene rings have
been, top and bottom, alternatively fused on (see
Fig.\ \ref{fig:polymers}(b)).  A reduced unit cell can be defined for
this system between the dashed lines in Fig.\ \ref{fig:polymers}(b),
with two 7-sites reduced unit cells per unit cell.  It is a half-filled
band system and, like PAA, a zero-width band gap is predicted.
Resonance theory, following Ref.\ \cite{Klein86}, predicts two
maximum-cardinality degenerate Kekul\'e phases.  As in the PAA
polymer, the interesting distortions that could open the band gap
and lift the degeneracy are those that are {\it antisymmetric\/}
under operations which interchange the two types of reduced unit
cells.

The PPR polymer is formed by fused benzene rings as drawn in
Fig.\ \ref{fig:polymers}(c).  The unit cell containing 10 sites
is defined between the dashed lines in the graph, and there is no
smaller reduced unit cell for this system.  The space group is
generated by the point group $D_{2h}$ and the translation
operations along the strip (see Table \ref{tab:grups} and
Fig.\ \ref{fig:symmetries}(c)).  Differently from the rest
of the polymers here, there is no glide-reflection symmetry
operation for PPR.  Furthermore, it does not have an odd number
of $\pi$-electrons per reduced unit cell so that it does not
correspond to a half-filled band system.  Nevertheless,
there is an accidental degeneracy at the H\"uckel level
of approximation, so that it has a zero-width band
gap anyway (see next section).  Correspondingly, resonance theory
predicts two maximum-cardinality degenerate Kekul\'e phases
\cite{Klein86}.  A {\it totally symmetric\/} distortion will also
be considered for this system (see Table \ref{tab:PPR}).

\section{Translationally adapted H\"uckel model}\label{sec:huckel}

The H\"uckel model is the simplest tight-binding model:
\begin{equation}
H_{Huck} = \sum_{<ni,mj>,\sigma} \beta_{ni,mj} \left( c_{ni
\sigma}^+ c_{mj \sigma} + c_{mj \sigma}^+ c_{ni \sigma} \right)
\label{eqn:Huck}
\end{equation}
$c_{ni \sigma}^+$ ($c_{ni \sigma}$) are the creation (annihilation)
electron operators on site $i$ of unit cell $n$ with spin
$\sigma$ and $\beta_{ni,mj}$ is the ``H\"uckel resonance integral''
(or hopping integral) between sites $i$ and $j$ in unit
cells $n$ and $m$, respectively.  $<ni,mj>$ indicates that the sum
is restricted to nearest neighbors.  Considering the translational
invariance symmetry of the system, we can define translationally
symmetry adapted states
\begin{equation}
\mid j;k \rangle \equiv \frac{1}{\sqrt{L}} \sum_{n=1}^{L}
e^{ikn} \mid n,j\rangle, \quad k= \frac{2\pi n_k}{L}, \quad n_k =
0,1, \cdots ,L-1.
\end{equation}
The matrix elements of the Hamiltonian between these new states are
\begin{equation}
\langle j;k \mid H \mid i;k'\rangle = \delta_{kk'} \sum_{<ni,mj>}
e^{-ik (n-m)} \beta_{ni,mj}.
\label{eqn:H}
\end{equation}
Diagonalizing the hamiltonian matrix elements, the energy bands
$\varepsilon(k)$ are finally obtained.

Symmetry breaking can be considered taking $\beta_{ni,mj}$ as
\begin{equation}
\beta_{ni,mj} = \beta (1 + \Delta_{ni,mj})
\end{equation}
where $\Delta_{ni,mj}$ ($\mid\Delta_{ni,mj}\mid \ll 1$) is the
distortion parameter, as introduced in the previous section, that
measures the strength of the distortion between sites $ni$ and
$mj$.

\section{Valence Bond method}\label{sec:VB}

Within the VB picture we attempt here to go beyond resonance theory
when solving the Heisenberg Hamiltonian:
\begin{equation}
H_{Heis} = \sum_{<ni,mj>}J_{ni,mj}\vec{S}_{ni} \vec{S}_{mj}
\end{equation}
$J_{ni,mj}$ is the ``exchange integral'' between nearest-neighbor
sites $ni$ and $mj$ and $\vec{S}_{ni}$ denotes the spin operator
for one electron on site $ni$.
\begin{equation}
J_{ni,mj}= J(1+ \Delta_{ni,mj})
\end{equation}
with $\Delta_{ni,mj}$ being the distortion parameter associated to
the bond between sites $ni$ and $mj$ when there is a symmetry
breaking.

While solving the H\"uckel model is an easy task, solving the
Heisenberg hamiltonian is in general a non-trivial problem.  In
order to obtain, along with the appropriate approximate
wave-functions, good variational upper bounds to the ground-state
energy of this model, $E(\Delta)$, for the polymer systems, we
consider two different types of cluster-expanded {\it ans\"atze\/}
that depend on variational parameters, each of which describes the
{\it local\/} features of the system.  Since our polymers are
bipartite systems with total spin zero, we have considered a {\it
N\'eel-state-based ansatz\/} and a {\it RVB ansatz\/}.  These {\it
ans\"atze\/} were introduced in Ref.\ \cite{pace} and we shall make
here a brief description of them.  Related {\it ans\"atze\/} have
also been successfully considered by other authors
\cite{Doucot,Sachdev} when solving the $s=1/2$ Heisenberg
hamiltonian for the square lattice.

\subsection{N\'{e}el-state-based ansatz (NSBA)}

The cluster expanded wave-function {\it ansatz\/} in this section
is based upon the N\'{e}el state as a zeroth-order wave-function,
\begin{equation}
\mid \Phi_N \rangle = \prod_i^{i\in A}\alpha(i)\prod_j^{j\in
B}\beta(j)
\label{eqn:ne}
\end{equation}
where $A$ and $B$ denote the two sets of sites such that each
member of one set is a nearest neighbor solely to (some) sites of
the other set, and $\alpha(i)$ ($\beta(i)$) indicate that the spin
of the electron on site $i$ is $+1/2$ ($-1/2$).  A lowering of the
energy, with respect to that of the N\'eel state, occurs for an
{\it ansatz\/} defined within a subspace spanned by
$\mid\Phi_N\rangle$ and the states obtained when
applying to $\mid\Phi_N\rangle$ the XY terms,
$S_{ni}^{\pm}S_{mj}^{\mp}$, of the Heisenberg operator, an
arbitrary number of times in an ``unlinked'' way.  These additional
states which are to be mixed with the N\'eel state can be generated
in terms of the nearest-neighbor pair excitation operator:
\begin{equation}
P\equiv \sum_{ni}^{\in A}\sum_{mj}^{<ni,mj>}
x_{ni,mj}S_{ni}^{-}S_{mj}^{+}
\label{eqn:pex}
\end{equation}
where the $x_{ni,mj}$ are scalars to be optimized and $S_{ni}^{-}$
and $S_{mi}^{+}$ are spin raising and lowering operators on site
$ni$
\begin{equation}
S_{ni}^{\pm} \equiv S_{ni}^x \pm {\rm i} S_{ni}^y
\end{equation}

From that, the N\'eel-state-based wave-function {\it ansatz\/}
(NSBA) will be a cluster-expansion in terms of $P$ excitations
acting on the N\'eel state
\begin{equation}
\mid \Psi_N\rangle = U e^{P} \mid \Phi_N\rangle
\label{eqn:nsb}
\end{equation}
where $U$ indicates that only unlinked terms are to be retained
from the Taylor series expansion.  Namely, $\mid \Psi_N\rangle$ is
a wave-function where the N\'eel state is mixed with states that
differ from it by an arbitrary number of couples of disjoint pairs
of neighboring spins that have been flipped, each state in the
superposition being weighted by the product of the variational
parameters associated to the flips in that state.

\subsection{Resonating Valence Bond ans\"{a}tze}

In this approach we start with a one-bond-range RVB (1BR-RVB), that
plays the fundamental zeroth-order role for the more elaborated
three-bond-range RVB (3BR-RVB) {\it ansatz\/} following.

\subsubsection{One-Bond-Range RVB ansatz}

A 1BR-RVB $\mid\Psi_1\rangle$ is a weighted superposition of
Kekul\'{e} states, i.e.\ nearest neighbor VB states, where every
site $ni$ is spin paired to one of its neighbors $mj$.  It can be
written as
\begin{equation}
\mid \Psi_1\rangle = U_0 \prod_{ni}^{\in A}\sum_{mj}^{<ni,mj>}
x_{ni,mj} \left(I-S_{ni}^- S_{mj}^+\right) \mid \Phi_N\rangle
\label{eqn:srvb}
\end{equation}
$I$ is the identity operator, $U_0$ indicates that the terms to be
retained are those where each site appears once and only once, and
the weighting factor of a Kekul\'{e} state in $\Psi_1$
is a product of variational parameters $x_{ni,mj}$ associated to
the singlet pairs $ni,mj$ in the Kekul\'{e} state considered.

\subsubsection{Three-Bond-Range RVB ansatz}

The 3BR-RVB is a weighted superposition of all the VB structures
within a phase with each spin-pairing between A- and B-sublattice
sites separated by no more than 3 bonds.  In the usual form for
cluster expanded wave-functions, it may be viewed as generated from
the 1BR-RVB as follows:  The XY terms, $S_{ni}^{\pm}
S_{mj}^{\mp}$, of the Heisenberg Hamiltonian acting on the 1BR-RVB
wave-function of Eqn.\ (\ref{eqn:srvb}) yield ``long-bonded''
states with pairings among three-bond distant neighbors, along with
``neighbor-bonded'' states already in $\Psi_1$.  These
``long-bonded'' states can be directly generated by the
``recoupling'' of two neighboring bond-singlets in $\Psi_1$ (see
Ref.\ \cite{pace}).  We may denote by $\hat{q}_{ef}$ the operator
related to such a recoupling between two bond-singlets $e$ and $f$.
From $\mid \Psi_1\rangle$ we may build the 3BR-RVB allowing an
arbitrary number of recouplings of two simply neighboring
bond-singlets, i.e.\ unlinked pairs with one and only one site in a
pair being a nearest neighbor to a site in the other pair.  Then,
the overall 3BR-RVB excitation operator above the 1BR-RVB
wave-function might be viewed to be
\begin{equation}
Q=\sum _{< e,f >} x_{ef}\hat{q}_{ef}
\end{equation}
with the $x_{ef}$ being variational parameters, and where $< e,f >$
indicates that the sum is restricted to simply neighboring
bond-singlets.  The corresponding {\it ansatz\/} would then be
\begin{equation}
\mid \Psi_3 \rangle = U e^Q \mid \Psi_1\rangle
\label{eqn:lrvb}
\end{equation}
where again $U$ indicates that only unlinked terms are to be
retained.  That is, in the Taylor series expansion of $e^{Q}$ one
retains only products of $\hat{q}_{ef}$ such that no pair index
($e$ or $f$) shares any vertices with another pair index in the
product.  And $Q$ and $\Psi_1$ are to be optimized simultaneously.

\subsection{Expectation-value calculations by the Transfer-Matrix
Technique}\label{sec:T-M}

The ground-state energy
\begin{equation}
E(\Psi) = \frac{\langle\Psi | H | \Psi\rangle}{\langle\Psi |
\Psi\rangle}
\label{eqn:energy}
\end{equation}
is computed as a function of variational parameters for each
of the above introduced {\it ans\"atze\/} assuming translational
symmetry and cyclic boundary conditions along $L$.  The way our
{\it ans\"{a}tze\/} are chosen allows us to deal with the systems
locally, so that one can define a transfer matrix \cite{pace}
that describes the local features and reduces the computation of
Eqn.\ (\ref{eqn:energy}) to products of ``small'' matrices
\cite{pa,pace,Klein86-2,Mod,teochem,croatica,Havilio}.  Let's
suppose there are imaginary vertical lines cutting the strip on
translationally equivalent positions (including improper
translations).  We can define the {\it ansatz\/}-dependent ``local
states'' according to every possible local spin-pairing/spin-flip
pattern around a given position determined by one of the imaginary
vertical lines, and ultimately use this to compute $\langle \Psi
\mid \Psi \rangle$.  Thence these local states contain the
contributions from both the {\it bra\/} and the {\it ket\/}.
From the assumed translational symmetry, local states in every
position are to be the same.  Now, labelling these local states by
$e_t$, $t$ ranging over the whole set of local states, we let the
transfer-matrix element
\begin{equation}
T_{ts} \equiv (e_t \mid T \mid e_s)
\label{eqn:T}
\end{equation}
denote a weighted sum over the various ways a local state $e_s$
may succeed a local state $e_t$.  The weight of every contribution
is obtained by considering the variational parameters associated
to the way $e_t$ evolves to $e_s$, and, eventually, additional
factors coming from Pauling's superposition rules \cite{Pauling-2}.
The overlap is then evaluated in terms of the $T$ matrix:
\begin{equation}
\langle \Psi \mid \Psi\rangle ={\rm tr} T^L
\label{eqn:overl}
\end{equation}
For $L \rightarrow \infty$, the largest eigenvalue $\Lambda$ of $T$
dominates, and the overlap reduces to
\begin{equation}
\langle\Psi \mid \Psi\rangle \simeq \Lambda^L
\end{equation}

The hamiltonian expectation value over $\mid \Psi \rangle$ can be
obtained in a similar way introducing a ``connection'' matrix, $C$,
defined according to
\begin{equation}
\langle \Psi \mid H \mid \Psi \rangle = JL \langle \Psi \mid
\sum_{<ni,mj>}^{\rm per\; cell} \vec{S}_{ni}
\vec{S}_{mj} \mid \Psi \rangle =JL\; {\rm tr} \left\{ T^{L-c} C
\right\}
\label{eqn:hamil}
\end{equation}
where $c$ measures the range of the interaction within the {\it
ansatz\/}.  In our case $c=2$.  And, the matrix element
\begin{equation}
C_{ts} = ( e_t \mid C \mid e_s )
\end{equation}
is a weighted sum over the various ways a local state $e_s$ may
succeed a local state $e_t$ after $c$ transfer-matrix-steps
when the Hamiltonian operators per unit cell are present.
In the long-length limit Eqn.\ (\ref{eqn:energy}) reduces to:
\begin{equation}
E =\frac{1}{ \Lambda ^2} {{(\Lambda ,l \mid C \mid \Lambda ,r)}
\over {(\Lambda,l \mid \Lambda ,r)}}
\label{eqn:ene2}
\end{equation}
where $(\Lambda , l\mid $ and $\mid \Lambda, r)$ are left and
right eigenvectors corresponding to the maximum eigenvalue
$\Lambda$ of $T$.  This expression is a function of the variational
parameters associated to $\Psi$ and an upper bound to the exact
ground-state energy is obtained.   Implementation of a suitable
numerical optimization yields a best upper bound.
The energy expression (\ref{eqn:ene2}) can be readily generalized
when considering possible distortions.  The connection matrix per
unit cell can be understood as a sum of matrices $C_{ni,mj}$, each
one concerning two body interactions between neighboring sites $ni$
and $mj$, weighted by the factor $1+\Delta_{ni,mj}$ that modifies
its interaction strength.  Then
\begin{equation}
C = \frac{1}{2} \sum_{i}^{\in n}
\sum_{mj}^{<ni,mj>}  \left( 1 + \Delta_{ni,mj}\right) C_{ni,mj}.
\end{equation}
and the energy expectation evaluation is reduced to some ``simple''
matrix manipulations.

\section{Results and Discussion}\label{sec:resultats}

Computations based on band theory at a H\"uckel tight-binding level
of approximation (see section \ref{sec:huckel}), and within VB
theory with the cluster expanded 1BR-RVB {\it ansatz\/} and NSBA
(see section \ref{sec:VB}) were carried out for all polymer systems
here described.  For the PAA also the 3BR-RVB has been used.  In
this case the 1BR-RVB {\it ansatz\/} contains only one Kekul\'e
structure, so it is especially appropriate to use the 3BR-RVB
wave-function and go beyond a single Kekul\'e-structure 
approximation.  This circumstance differs with the rest of the
polymers, where the number of Kekul\'e structures in the
corresponding 1BR-RVB is big.  The different VB upper bounds to
the energy of the undistorted polymers are presented in Table
\ref{tab:energ}, together with that for polyacetylene.  The
lowest upper bound to the ground-state energy for the undistorted
system is given by the NSBA.

\subsubsection{PAA}

The highest occupied H\"uckel tight-binding band and the
lowest unoccupied band cross at $k = \pi$.  Taking into account the
perturbation $\Delta_{ni,mj}$ in the H\"uckel resonance integral,
$\beta_{ni,mj} = \beta (1 + \Delta_{ni,mj})$, only the {\it
totally-antisymmetric\/} distortion with respect to $C_s$ and
$\sigma_v$ labelled as $C_3$ (see Table \ref{tab:PAA}) opens
a gap at $k = \pi$.  But the leading term of the energy
lowering $\bigtriangleup E$ versus $\Delta$ is $\sim \Delta^2$, as
it is the positive phonon energy contribution to be added.
So band theory at this low level of approximation predicts neither
a presence nor absence of a $C_3$ distortion for this system,
the result depending on the final balance between these two
contributions to the energy. Nevertheless, if interactions with
more distant $\pi$-centers are included, although small, linear
terms in $\Delta$ are argued to arise \cite{Klein86} and then the
distortion is favored.

Still within band theory, this system has also been studied by
other authors at different levels of approximation.  Kertesz
\cite{Kertesz} and Tanaka \cite{Tanaka-87} suggest a totally
antisymmetric distortion, though leading to a quadratic small gap
that could be suppressed by interchain interactions.  The
tight-binding SCF-MO method at the level of CNDO/2 (complete
neglect of differential overlap) calculations suggests that the
Peierls distortion does not take place so one can expect {\it
metallic\/} behavior \cite{Yamabe}, while Bozovi\'c \cite{Bozovic}
combining tight-binding calculations with group-theoretical
arguments predicts distortions of type $B$ (see Table
\ref{tab:PAA}) as favored.  Therefore, within band theory, 
predictions about the opening or not of a band gap at the Fermi
level, or the distortion driving it, depend crucially on the level
of approximation.

Let us consider now the many-body VB method.  The ground-state
energy has been obtained using the NSBA and both the 1BR- and the
3BR-RVB {\it ans\"atze\/} of section \ref{sec:VB}, as a function of
$\Delta$ for the different distortions $A, B_1, B_2, C_1, C_2, C_3$
(see Table \ref{tab:PAA}).  Transfer and connection matrices
of dimensions $14 \times 14$ (for the NSBA) and $60 \times
60$ (for the 3BR-RVB {\it ansatz\/}) were needed in order to
carry out computations.  The energy for the different distortions
when the NSBA is used has been plotted as a function of $\Delta$ in
Fig.\ \ref{fig:EPAA}(a), while results obtained with the 3BR-RVB
{\it ansatz\/} are presented in Fig.\ \ref{fig:EPAA}(b).  Plots
from the 1BR-RVB {\it ansatz\/} are not given, since they are
qualitatively identical to those from the 3BR-RVB ones.  Comparing
NSBA and RVB {\it ans\"atze\/}, it can be seen that the ordering of
$\bigtriangleup E$ for the different distortions is the same in any
case, the strongest lowering corresponding to the $C_3$ distortion.

Nevertheless, while the energy response to $A$, $B$ and $C$
distortions is linear for the RVB {\it ans\"atze\/}, clearly predicting
a $C_3$ distorted ground-state, in the NSBA case they still go as
$\sim \Delta^2$.  Fitting the results in a parabolic curve, it is
obtained that $\bigtriangleup E \sim -1.923 \Delta^2$.  Again a
distortion is not clearly predicted with our NSBA.  A comparison
of the coefficients coming from this term and those from the phonon
energy should be made in order to decide whether this {\it ansatz} is
able to predict or not to predict a $C_3$ distortion.  This ambiguity
of prediction in some sense rationalizes earlier contradictory results:
via numerical band theory by Yamabe {\it et al.} \cite{Yamabe},
predicting an undistorted ground-state, and via band/group-theoretic
considerations by Bozovi\'c \cite{Bozovic}, predicting a $B$
distortion.

Although the RVB ground-state energy is higher than the NSBA, its
predictions on ground-state instabilities are based upon the known
global-singlet character of the ground state along with its
local-singlet character, leading to asymptotically orthogonal and
noninteracting phases responding essentially independently to
distortions.  Relaxation of this local-singlet character would imply
the inclusion of pairings between distant sites, leading to 
undesirable long-range correlations of the type in the N\'eel state.
Then we expect a $C_3$ distorted ground-state as predicted by RVB.
Furthermore, NSBA at this lower level, with only two-site excitations,
does not always seem sensitive to instabilities as at higher order,
such as for polyacetylene \cite{pa}, then we expect that the distortion
could be clearly predicted when going to a higher order NSBA too.
Also it can be argued that inclusion of slightly longer-range
interactions (as between next-nearest neighbors) in the hamiltonian
will increase the `frustration' and the NSBA energy whereas the
RVB expectations will change but little.  Thus there is a tendency
to invert the energy ordering of these states.  Still another
argument favoring RVB predictions is that the NSBA is not a pure
singlet, as the ground state is known to be.  Also the RVB type 
{\it ansatz} accords more closely to a classical organic chemical
view of these polymers.

\subsubsection{PBA}

The lowest occupied H\"uckel tight-binding band and the highest
unoccupied one cross at $k = \pi$, so it is a zero-width band gap
system.  From all the possible distortions considered in Table
\ref{tab:PBA}, only the {\it totally antisymmetric\/} distortions
$C_1$ and $C_2$ open a gap with an energy dependence linear in
$\Delta$.  Therefore, band theory predicts that the system will
distort.  In the VB picture the possible distortions in PBA have been
studied with the 1BR-RVB {\it ansatz\/}.  For this system we only
carried out calculations with this {\it ansatz\/} for two reasons i)
the 1BR-RVB {\it ansatz\/} gives already a good upper bound to the
ground-state energy because there is mixing of Kekul\'e states and
ii) the dimension of the transfer and connection matrices for the
3BR-RVB and the N\'eel-state-based {\it ans\"atze\/} grow
substantially with respect to the 1BR-RVB one.  In
Fig.\ \ref{fig:EPBA} the energy of the 1BR-RVB {\it ansatz\/} has
been plotted as a function of $\Delta$ for the distortions $A_1$,
$A_2$, $C_1$ and $C_2$ classified in Table \ref{tab:PBA}.  The most
favored distortions are the {\it totally-antisymmetric\/} ones
$C_1$ and $C_2$, in particular $C_1$ with a dependence $\sim
\Delta$.  This result agrees with the predictions given from band
theory, concluding that complementary approaches lead to the same
kind of distortions for this system.

\subsubsection{PPR}

PPR is not a half-filled band system but the H\"uckel model predicts
an accidental zero-width band gap at $k = 0$.  $A$, $B$, $C$ and
$D$ distortions (see Table \ref{tab:PPR}) have been considered.
The distortions $C_1$ and $C_2$ open a gap at $k = 0$ weakly,
with an energy dependence $\bigtriangleup E \sim \Delta^2$.  But
the {\it totally symmetric\/} distortion $D_1$ opens a band gap
with an energy response linear in $\Delta$.  This result agrees
with the predictions given by Bozovi\'c \cite{Bozovic} and Tanaka
{\it et al.} \cite{Tanaka-PPR}.  In Fig.\ \ref{fig:EPPR}(a) the
N\'eel-state-based energy obtained, using $5 \times 5$ transfer and
connection matrices, is plotted as a function of $\Delta$ for
various possible distortions (see Table \ref{tab:PPR}).  Clearly
the {\it totally symmetric\/} distortion, $D_1$ is favored with a
linear energy dependence on $\Delta$.  Also in
Fig.\ \ref{fig:EPPR}(b) the 1BR-RVB energy is plotted for the
various distortions as a function of $\Delta$ and results agree with
the NSBA energy, namely that the $D_1$ distortion is the most favored
one with a linear dependence in $\Delta$.  As in PBA, the 1BR-RVB
{\it ansatz\/} gives already a good upper bound due to the mixing
of Kekul\'e states.

Band theory and the many-body VB method predict the same 
distortional behavior for this system, i.e.\ the system is
unstable to a {\it totally symmetric\/} $D_1$ distortion.
Some evidence exists for polyperylene synthesis \cite{Murakami},
but further experimental information on the structure (and
properties) of this system is still needed.

\section{Conclusions}\label{sec:conclusions}

We have presented, both with the simple H\"uckel tight-binding
band theory and with a Heisenberg model hamiltonian (or,
equivalently, the VB model), a study of the ground-state nature
of a family of polymers:  polyaceacene, poly(benz[m,n]anthracene),
and polyperylene.  We have focused our attention
on correspondences between Peierls and spin-Peierls instabilities
predictions, when analyzed from these two complementary approaches.

Upper bounds to the energy of the Heisenberg model in each case
have been obtained with two alternative localized-site
cluster-expanded wave-functions, i.e.\ RVB-type {\it ans\"{a}tze\/}
and a N\'{e}el-based {\it ansatz\/}.  We have shown that simple
expressions of the physical magnitudes we were interested in, were
easily obtained by using the transfer-matrix technique of
ref.\ \cite{pace}.

From our results, it is concluded that the RVB wave-functions
considered, which are restricted to 1BR type for all the systems
other than PAA, do not give our best upper bound to the
ground-state energy of the undistorted systems.  Nevertheless, they
are relevant for studying such phenomena as the {\it spin-Peierls
instability\/} and elementary excitations as hole excitations or
excitonic excitations as already pointed out \cite{teochem}.
Moreover the RVB {\it ans\"atze\/} have a global-singlet character and a
local-singlet character, precluding long-range order of the type of
the N\'eel state, and generally improve relative to N\'eel-based {\it
ans\"atze\/} upon inclusion of higher-order (frustrative) terms in an
elaborated Heisenberg model.

The N\'{e}el-state-based {\it ansatz\/} gives a fairly good upper
bound to the ground-state energy for all the systems considered.
For the nearest-neighbor model considered, this {\it ansatz\/}
always yields lower energy than the RVB ones for undistorted
systems.  We have shown that with such a simple
N\'{e}el-state-based wave-function, the corresponding energy is
notably lower than the energy of the N\'{e}el state, while
computations remain fairly simple.  The N\'{e}el-state-based {\it
ansatz\/} predicts for the polymers studied, the same distortions
as the RVB description, except for the case of polyaceacene where
this {\it ansatz\/} in our current simple considerations does not
show whether the distortion is going to take place or not, although
the strongest lowering of the energy also correspond a {\it totally
antisymmetric\/} distortion.

From the Heisenberg hamiltonian, or equivalently from the VB model,
we have obtained that:
\begin{enumerate}
\item  PAA shows a {\it totally antisymmetric\/} distortion from the
RVB, while NSBA is not conclusive, depending on the balance between
electronic energy lowering and the phonon energy contribution.
\item  PBA shows a {\it totally antisymmetric\/} distortion.
\item  PPR is unstable to a {\it totally symmetric\/} distortion.
\end{enumerate}
Within the band-theoretic picture, the H\"uckel tight-binding model
has been studied for all the same polymers.  Results obtained for
our $\pi$-network system are as follows:
\begin{enumerate}
\item  PAA could show a {\it totally antisymmetric\/} distortion at
a simple H\"uckel level, depending on the balance between
electronic energy lowering and the phonon energy contribution.
Other approximations already in the literature
\cite{Bozovic,Yamabe} yield contradictory results.
\item  PBA shows a {\it totally antisymmetric\/} distortion.
\item  PPR shows a {\it totally symmetric\/} distortion.
\end{enumerate}

Comparing band theory and the Heisenberg model results, it can be
concluded that predictions of these two models based on opposite
(or complementary) limits seem to lead to similar consequences
under similar structural circumstances, i.e.\ both approaches
predict the presence or absence of the same instability to
symmetry for the polymers.  It is to be noted that the
band-theoretic results depend crucially on the level of
approximation, as it is observed in the study of polyaceacene,
where this picture at different levels of approximation gives rise to
different predictions.  On the other hand the Heisenberg model has
proven to give predictions consistent from one level to another.
Even in the case of PAA, where NSBA cannot make a clear prediction as
it happens with band theory at its lower level, NSBA still shows the
strongest lowering of the energy for the very same distortion
suggested by RVB.\@ Since the NSBA at this lower level, with only
two-site excitations, does not always seems so sensitive to
instabilities as at higher orders, such as it happens with
polyacetylene \cite{pa}.  That is, the distortion sometimes 
seems to only occur with a higher order NSBA, in agreement
with RVB results.  Therefore, it seems that the VB model,
which includes correlation explicitly, gives a good description
of these benzenoid systems, predicting spin-Peierls distortions
whenever a Peierls distortion is also predicted.  These results
modify earlier suggestions (see ref.\ \cite{dixit}) that
inclusion of correlation ``a posteriori'', as a perturbation,
diminishes the distortion.  That is, we find any diminishment
does not go to zero in the (strong correlation)
Heisenberg-model limit, and indeed the RVB results indicate a
stronger response to distortions (at least at the undistorted
point on the potential energy hypersurface).

It has been shown that this treatment is computationally feasible
specially for quasi-one dimensional systems where the
transfer-matrix technique proves to be a powerful tool of
computation.  It is important to note that the results are
developed in terms of quantities which remain finite as the
strip-length goes to infinity.  It is of some interest to compare
the computational effort involved in the tight-binding approach
versus that involved in our transfer-matrix cluster-expansion
approach (for either RVB or N\'eel-state-based wavefunctions).
The matrices ${\bf H}(k)$ of eqn.\ \ref{eqn:H} and ${\bf T}$ of
eqn.\ \ref{eqn:T} arise in these respective approaches and are both
finite independently of $L \rightarrow \infty$.  Both types are to
be diagonalized, but there are some differences:
\begin{enumerate}
\item  Typically ${\bf T}$ increases in size much more rapidly with
unit-cell ``width'' than does ${\bf H}(k)$ (though these behaviours
are reversed if the unit-cell ``length'' is considered instead).
\item  The total energy requires sampling many of the $L \rightarrow
\infty$ ${\bf H}(k)$ matrices (varying smoothly with wavevector
$k$) whereas for given parameters there is but one ${\bf T}$
matrix to treat.
\item  The optimal total energy for the cluster expansions entails
treating ${\bf T}$ matrices for numerous variational parameter
values whereas there is no much repetition with the ${\bf H}(k)$.
\end{enumerate}
Notably if one goes beyond the tight-binding method to Hartree-Fock
(or density-functional) approaches this last noted difference no
longer occurs.  Evidently the computational effort via either SCF or
our cluster expansion is roughly comparable (at least for linear
polymers with modestly sized unit cells).

The analysis carried out in this paper would require experimental
testing.  Though the synthesis of some of the systems considered
like PAA seems quite difficult to achieve, there are hopes in this
direction.  Finally, some aspects of this treatment are not
restricted only to the model Hamiltonian and the ground-state {\it
ans\"atze\/} presented but can be applied to any system with
effective short-range interactions if described by a localized-site
cluster expanded ground-state wave-function.

\acknowledgements

The authors acknowledge the University of Barcelona for computer
facilities and CPU time.  One of us (MAGB) acknowledges the DGICYT
(project PB92-0868).  RV acknowledges support through the Deutsche
For\-schungs\-gemein\-schaft.  DJK acknowledges support to the
Welch Foundation of Houston (Texas).

\newpage

\begin{table}
\protect\caption{Number of sites in the unit cell (uc) and in the
reduced unit cell (ruc), and symmetry operations in the space
group not including primitive translations.\label{tab:grups}}
\begin{center}
\begin{tabular}{|c||c|c|c|}
Polymer & sites in uc & sites in ruc & symmetries \\
\hline \hline
\rule[-3mm]{0mm}{8mm}
PAA & 6 & 3 & $i, \sigma_h, C_{2a}, C_{2b}, \sigma_v, C_s$ \\
PBA & 14 & 7 & $i, \sigma_h, C_2, C_{sa}, C_{sb}$ \\
PPR & 10 & -- & $i, \sigma_h, C_{2a}, C_{2b}, C_{2c},
\sigma_{v1}, \sigma_{v2}$
\end{tabular}
\end{center}
\end{table}

\begin{table}
\caption{Distortions considered for the PAA strip.  For $B$
distortions we identify subcases:  $B_1$ for $\Delta_1 > 0$ 
and $\Delta_2 = 0$; and $B_2$ for $\Delta_1 = 0$ and $\Delta_2 > 0$.
For $C$ distortions we identify subcases:  $C_1$ for $\Delta_1 > 0$ 
and $\Delta_0 = 0$; $C_2$ for $\Delta_1 = 0$ and  $\Delta_0 > 0$;
and $C_3$ for $\Delta_1 > 0$ and $\Delta_0 < 0$.\protect\label{tab:PAA}}
\begin{center}
\begin{tabular}{|c||c|c|c|}
Distortion & $C_s$ & $\sigma_v$ & Restrictions\ on\ $\Delta_l$ \\
\hline\hline
\rule[-3mm]{0mm}{8mm}
$A$ & +1 & -1 &
\begin{tabular}{c} $\Delta_0 = \Delta_{\bar{0}}
  = \Delta_2 = \Delta_{\bar{2}} = 0 $\\
 $ \Delta_1 = \Delta_{\bar{1}} =- \Delta_{1'} =
 - \Delta_{\bar{1}'}$
\end{tabular}\\ \hline
\rule[-3mm]{0mm}{8mm}
$ B$ & -1 & +1 &
\begin{tabular}{c}$ \Delta_0 = \Delta_{\bar{0}}
 = 0 $\\
 $\Delta_1 = -\Delta_{\bar{1}} =
 -\Delta_{1'} = \Delta_{\bar{1}'}$ \\
 $\Delta_2 = -\Delta_{\bar{2}}$
\end{tabular} \\ \hline
\rule[-3mm]{0mm}{8mm}
$ C$ & -1 & -1 &
\begin{tabular}{c}$ \Delta_1 = -\Delta_{\bar{1}} = \Delta_{1'} =
-\Delta_{\bar{1}'}$ \\
$\Delta_2 = \Delta_{\bar{2}} = 0 $\\
$\Delta_0 = -\Delta_{\bar{0}}$
\end{tabular}
\end{tabular}
\end{center}
\end{table}

\begin{table}
\caption{Distortions considered for the PBA strip.  All possible
$\Delta_i$,  $i= 1,2,3,4$ are assumed to be mutually independent.
For $A$ distortions we identify subcases:  $A_1$ for $\Delta_1 > 0$
and $\Delta_2 = \Delta_3 = \Delta_4 = 0$;  and $A_2$  for
$\Delta_1 = \Delta_2 = 0$, $\Delta_3 > 0$ and $\Delta_4 < 0$.
For $C$ distortions we identify subcases:  $C_1$ for $\Delta_1 > 0$
and the rest equal to zero;  and $C_2$ for $\Delta_1 = \Delta_2 = 0$,
$\Delta_3 > 0$ and $\Delta_4 < 0$.\protect\label{tab:PBA}}
\begin{center}
\begin{tabular}{|c||c|c|c|}
\rule[-3mm]{0mm}{8mm}
 Distortion & $C_s$ & $\sigma_v$ & Restrictions\ on\  $\Delta_l$ \\
\hline\hline
\rule[-3mm]{0mm}{8mm}
 $ A$ & +1 & -1 &   $\Delta_i = \Delta_{\bar{i}}
  = -\Delta_{i'} = -\Delta_{\bar{i}'} $\\  \hline
\rule[-3mm]{0mm}{8mm}
 $B$ & -1 & +1 &   $\Delta_i = -\Delta_{\bar{i}}
 = -\Delta_{i'} =  \Delta_{\bar{i}'} $
  \\ \hline
\rule[-3mm]{0mm}{8mm}
 $C$ & -1 & -1 &    $\Delta_i = -\Delta_{\bar{i}} =
   \Delta_{i'} = -\Delta_{\bar{i}'}$
 \end{tabular}
\end{center}
\end{table}
\newpage
\begin{table}
\caption{Distortions considered for the PPR strip, where  $j =
1,2,3$ and $i = 1,2,3,4$.  For $C$ distortions we identify subcases:
$C_1$ for $\Delta_1 = \Delta_4 = 0$, $\Delta_2 > 0$ and $\Delta_3 >
0$; and $C_2$ for $\Delta_1 = \Delta_4 = 0$, $\Delta_2 < 0$ and
$\Delta_3 > 0$.  For $D$ distortions we identify subcase $D_1$ for
$\Delta_1 =\Delta_2 = 0$, $\Delta_3 > 0$ and $\Delta_4 >
0$.\protect\label{tab:PPR}}
\begin{center}
\begin{tabular}{|c||c|c|c|}
\rule[-3mm]{0mm}{8mm}
 Distortion & $C_2$ & $\sigma_v$ & Restrictions\ on\  $\Delta_l$ \\
\hline\hline
\rule[-3mm]{0mm}{8mm}
 $ A$ & +1 & -1 &  \begin{tabular}{c} $\Delta_j = \Delta_{\bar{j}}
  = -\Delta_{j'} = -\Delta_{\bar{j'}} $ \\
  $\Delta_4 =  \Delta_{4'} = 0$
  \end{tabular}\\ \hline
\rule[-3mm]{0mm}{8mm}
 $B$ & -1 & +1 &  \begin{tabular}{c} $\Delta_j = \Delta_{\bar{j}}
 = -\Delta_{j'} = \Delta_{\bar{j'}}  $ \\
  $\Delta_4 =  \Delta_{4'} = 0$
  \end{tabular} \\ \hline
\rule[-3mm]{0mm}{8mm}
 $C $& -1 & -1 &    $\Delta_i = - \Delta_{\bar{i}} =
   \Delta_{i'} = -\Delta_{\bar{i}'} $\\
  \hline
\rule[-3mm]{0mm}{8mm}
 $D $& +1 & +1 &  \begin{tabular}{c}  $\Delta_i = \Delta_{\bar{i}}
= \Delta_{i'} = \Delta_{\bar{i}'} $
 \end{tabular}
 \end{tabular}
\end{center}
\end{table}

\begin{table}
\protect\caption{Ground-state Heisenberg energy per site in $J$
units for the family of $\pi$-network polymers studied.  PA
stands for polyacetylene.  The first row corresponds to the
energy obtained with a single Kekul\'e structure $\mid K \rangle$.
$\mid\Psi_1\rangle$ stands for the 1BR-RVB \protect{\it ansatz\/}
of Eqn.\ (\protect\ref{eqn:srvb}), $\mid\Psi_3\rangle$ is the
3BR-RVB \protect{\it ansatz\/} of Eqn.\ (\protect\ref{eqn:lrvb}),
$\mid\Phi_N \rangle$ is the N\'eel state, and $\mid\Psi_N\rangle$
the N\'eel-state-based \protect{\it ansatz\/} of
Eqn.\ (\protect\ref{eqn:nsb}).  The last row corresponds to
the exact ground-state energy which is known only for the 1D case.
\protect\label{tab:energ}}
\begin{center}
\begin{tabular}{|l||c|c|c|c|}
 E/JN & PA & PAA  & PBA & PPR \\
\hline\hline
\rule[-3mm]{0mm}{8mm}
$\mid K\rangle$ & -.37500 & -.37500 & -.37500 & -.37500 \\
\hline \rule[-3mm]{0mm}{8mm}
$\mid\Psi_1\rangle$ & -.37500 & -.37500 & -.4339(3) & -.4435(2)\\
\hline \rule[-3mm]{0mm}{8mm}
$\mid \Psi_3\rangle$ & -.41100 & -.4539(5) &    -   &     -    \\
\hline \rule[-3mm]{0mm}{8mm}
$\mid \Phi_N\rangle$ & -.25000 & -.333(3) & -.3214(3) & -.32500 \\
\hline \rule[-3mm]{0mm}{8mm}
$\mid \Psi_N\rangle$ & -.4279(1) & -.4941(0) &   -   & -.4906(2) \\
\hline \rule[-3mm]{0mm}{8mm}
 exact               & -.4431(5)    &      -    &  -  &   - 
 \end{tabular}
\end{center}
\end{table}

\begin{figure}
\caption{Polymer systems.  Fragments of: (a) Poly\-ace\-acene (PAA).
(b) Poly\-(benz[m,n])\-anthracene (PBA).  (c) Poly\-peryl\-ene (PPR).
The region between the vertical dashed lines defines the unit cell of
PPR, while for PAA and PBA the reduced unit cell is
instead identified.}
\label{fig:polymers}
\end{figure}
\begin{figure}
\caption{Symmetry elements for:  (a) PAA.  (b) PBA.  (c) PPR.}
\label{fig:symmetries}
\end{figure}
\begin{figure}
\caption{Representation of the different non-mixing Kekul\'e phases
of PAA, each one containing essentially one Kekul\'e structure.}
\label{fig:phases}
\end{figure}
\begin{figure}
\caption{PAA analysis. (a)  Unit cell and reduced unit cell.
(b) Labels associated to bonds.  (c) Symmetry elements chosen to
label distortions:  the screw axis $C_s$ and the vertical plane
$\sigma_v$.}
\label{fig:PAA}
\end{figure}
\begin{figure}
\caption{PBA analysis. (a) Unit cell and reduced unit cell.
(b) Bond labels.  (c) Symmetry elements chosen to label
distortions:  a screw axis $C_s$ and a vertical plane $\sigma_v$.}
\label{fig:PBA}
\end{figure}
\begin{figure}
\caption{PPR analysis. (a) Unit cell.  (b) Bond labels.  (c)
Symmetry elements chosen to label distortions:  a two-fold rotation
axis $C_2$ perpendicular to the molecular plane, and a vertical
plane $\sigma_v$.}
\label{fig:PPR}
\end{figure}
\begin{figure}
\caption{Energy as a function of the distortion parameter $\Delta$
in PAA (a) when the N\'eel-state-based {\it ansatz\/}
$\mid\Psi_N\rangle$ is considered, (b) when the 3BR-RVB {\it
ansatz\/} $\mid\Psi_3\rangle$ is considered.  The curves correspond
to the different distortions given in Table \protect\ref{tab:PAA}:
($\Box$) $B_1$, ($\triangle$) $B_2$, ($\Diamond$) $C_2$,
($\circ$) $A$, ($\bullet$) $C_1$, (\protect\rule{2mm}{2mm}) $C_3$.}
\label{fig:EPAA}
\end{figure}
\begin{figure}
\caption{Energy as a function of the distortion parameter $\Delta$
in PBA when the 1BR-RVB {\it ansatz\/} $\mid\Psi_1\rangle$ is
considered.  The curves correspond to the different distortions
given in Table \protect\ref{tab:PBA}: ($\circ$) $A_1$, ($\Box$)
$A_2$, ($\triangle$) $C_2$, ($\Diamond$) $C_1$.}
\label{fig:EPBA}
\end{figure}
\begin{figure}
\caption{Energy as a function of the distortion parameter $\Delta$
in PPR:  (a) when the N\'eel-state-based {\it ansatz\/}
$\mid\Psi_N\rangle$ is considered; (b) when the 1BR-RVB {\it
ansatz\/} $\mid\Psi_1\rangle$ is considered.  The curves correspond
to the different distortions given in Table \protect\ref{tab:PPR}:
($\triangle$) $C_1$, ($\Box$) $C_2$, ($\Diamond$) $D_1$.}
\label{fig:EPPR}
\end{figure}

\end{document}